\documentclass[12pt]{iopart}   %% preprint
\usepackage{graphicx}
\usepackage{dcolumn}
\usepackage{bm}
\begin{document}

\title{Spectral Density of the Two-Impurity Anderson Model}

\author{Satoshi Nishimoto\dag, Thomas Pruschke\dag, Reinhard M.~Noack\ddag}

\address{\dag\ Institut f\"ur Theoretische Physik, Universit\"at 
G\"ottingen, D-37077 G\"ottingen, Germany}

\address{\ddag\ Fachbereich Physik, Philipps-Universit\"at Marburg,
D-35032 Marburg, Germany}

\ead{nishimoto@theorie.physik.uni-goettingen.de}

\begin{abstract}
We investigate static and dynamical ground-state properties of the two-impurity
Anderson model at half filling in the limit of vanishing impurity separation using the
dynamical density-matrix renormalization group method. In the
weak-coupling regime, we find a quantum phase transition as function
of inter-impurity hopping driven by the charge degrees of freedom. For
large values of the local Coulomb repulsion, the transition is driven
instead by a competition between local and non-local magnetic
correlations. 
We find evidence that, in contrast to the usual
phenomenological picture, it seems to be the bare effective exchange
interactions which trigger the observed transition. 
\end{abstract}
\maketitle

\section{Introduction}

Although forty years have passed since the discovery of the Kondo
effect, it is still one of the most interesting topics in condensed matter physics; 
it lies at the heart of understanding strongly correlated electron 
systems~\cite{hewson}. The Kondo effect, which leads to the quenching of an 
impurity spin, forms the basis of the physics of a single magnetic impurity 
embedded in a metal. However, systems with more than one impurity are 
considerably more complicated and present additional difficulties in a theoretical 
investigation. In particular, there are two effects which compete against each 
other in multiple-impurity systems: the Kondo effect and the 
Runderman--Kittel--Kasuya--Yosida (RKKY) interaction.
The RKKY exchange favors the formation of non-local magnetic correlations; 
the Kondo effect, on the other hand, is based on purely local magnetic correlations. 
The competition between Kondo effect and RKKY interaction is thought to be 
the key mechanism to understanding the magnetic properties of the heavy fermion 
materials~\cite{doniach}. The simplest systems in which to study this competition 
are two-impurity models. Recently, such models have also attracted much attention 
in the context of double quantum dots~\cite{jeong}, which can be viewed as a 
direct experimental realization of the two-impurity Kondo model.

Theoretically, the two-impurity Anderson model (TIAM)~\cite
{tsay,jayaprakash,chakravarty,jones,saso,andreani,schiller,santoro,klein,buesser,ivanov,neto,aguado}
and the two-impurity Kondo model
(TIKM)~\cite{jayaprakash2,fye,jones2,affleck,ingersent,sakai,sire,andreani2,gan,georges,silva,hallberg,schlottmann,izumida,aono,vojta,allub,campo}
have been extensively studied with various methods; nevertheless, their physical  properties at low temperature are not yet well understood. In particular, the 
situation is far from clear concerning dynamical properties because only a few 
methods are able to reliably calculate the dynamical properties of such models due 
to their complexity. So far, the spectral density has been calculated using 
perturbation theory (PT)~\cite{santoro,ivanov} and the numerical renormalization 
group (NRG)~\cite{sakai}, but the results are not fully  satisfactory. The PT 
provides an explicatory and accurate picture of quantum impurity dynamics only in 
certain limiting cases. While the NRG can determine the low-energy dynamics of 
quantum impurity models almost exactly, it is less precise at high energy.

Recently, the dynamical density-matrix renormalization group (DDMRG) 
method~\cite{eric} was applied to the single-impurity Anderson model (SIAM) 
and it was shown that the method can calculate the impurity spectral density 
with good resolution for all frequencies and coupling strengths~\cite{nishim,raas}. 
This method can be extended to investigate the dynamics of a two-impurity problem 
without difficulty. Here we study the spectral density of the TIAM using the  DDMRG method. Since the parameter space of the TIAM is rather large, for 
simplicity we focus here on the limit of small inter-impurity distance. This 
simplification does not change the substance of the problem and is quite likely 
relevant for typical experimental situations, e.g., clusters of magnetic atoms on 
metal surfaces or multiple quantum-dot systems.

The aim of this paper is to demonstrate the efficiency of the DDMRG for 
the two-impurity system and to discuss the dynamical properties of the TIAM in the 
limit of zero impurity distance as the first step of a more general DDMRG study.
The organization of this paper is as follows. In section 2, the model Hamiltonian 
for the TIAM is introduced. In section 3, we transform the Hamiltonian for an 
efficient treatment with the (D)DMRG and define even- and odd-parity orbitals 
of coupled impurities. In section 4, we show the static and the dynamical 
properties calculated with the (D)DMRG method. 
The conclusion and discussion follow in section 5.

\section{Model}

The Hamiltonian for two impurities placed at ${\bf R}_i$ ($i=1,2$) is written as 
\begin{eqnarray}
\nonumber
\hat{H} &=& \sum_{{\bf k}\sigma} \varepsilon_{\bf k} 
\hat{f}_{{\bf k}\sigma}^\dagger \hat{f}^{\phantom{\dagger}}_{{\bf k}\sigma}
+ \sum_{i{\bf k}\sigma} V_{\bf k} \left(e^{i{\bf k}\cdot{\bf R}_i} 
\hat{f}_{{\bf k}\sigma}^{\dag}\hat{d}^{\phantom{\dagger}}_{i\sigma} + h.c. \right) \\
\nonumber
&+& U \sum_{i=1,2} (\hat{n}^d_{i\uparrow} - \mu)
(\hat{n}^d_{i\downarrow} - \mu) 
+ t_{12} \sum_{\sigma} 
(\hat{d}_{1\sigma}^{\dag}\hat{d}^{\phantom{\dagger}}_{2\sigma} + h.c)
\label{hamiltonian}
\end{eqnarray}
where $\hat{d}_{i\sigma}^\dagger$ ($\hat{d}^{\phantom{\dagger}}_{i\sigma}$) creates (annihilates) an electron with spin $\sigma = \uparrow,\downarrow$ in a local level 
(the impurity site $i$), $\hat{n}^d_{i\sigma} = \hat{d}_{i\sigma}^\dagger \hat{d}^{\phantom{\dagger}}_{i\sigma}$and  
$\hat{f}^{\dag}_{\sigma} \ (\hat{f}^{\phantom{\dagger}}_{\sigma})$ 
creates (annihilates) an electron with spin $\sigma$ in an eigenstate 
of the (noninteracting) host band with dispersion $\varepsilon_{\bf k}$.
The sum over ${\bf k}$ runs over all states of the host band. 
The hybridization between the local impurity state and the delocalized
band state ${\bf k}$ is given by the positive couplings $V_{\bf k}$. 
Electrons in the local level are subject to a Coulomb repulsion $U$.
In this paper, the energy level of impurity sites is set by $\mu = -U/2$. 
Under this assumption, we can map our model to the two-impurity Kondo 
model in the strong-coupling limit. We also set ${\bf R}_1-{\bf R}_2={\bf 0}$ 
and replace ${\bf k} \to k$ for simplicity. Thus, our model depends on the 
parameter $U$ and the hybridization function
\begin{equation}
\Delta(\omega) = \pi \sum_{k} |V_{k}|^2 \delta(\omega - \varepsilon_{k}) 
\geq 0 \;  .
\label{hybridization}
\end{equation}
For a symmetric hybridization function, $\Delta(\omega)=\Delta(-\omega)$, the 
TIAM is particle-hole symmetric for $t_{12}=0$. 

Since, for the time being, we are interested in understanding the qualitative 
aspects of the model, it is convenient to choose a flat-band host density as 
the hybridization function. In addition, the flat-band case of the SIAM is very 
well understood~\cite{hewson} and is thus helpful to explain features found in
our model. We also take the host bandwidth to be much larger than any other 
bare energy scale and use a hybridization function which is constant, 
$\Delta(\omega) = \Delta_0$. Our goal is then to compute the spectral density in 
the relevant energy window $-W/2 < \omega < W/2$ with $W/2 > U/2, \Delta_0$. 
For all numerical results presented here, the energy scale is set by 
$\Delta_0 = 1/\pi$.

\section{Method}

In this work, we employ the DMRG technique~\cite{steve} which is a reliable 
numerical method for one-dimensional systems. We use the standard DMRG 
method to calculate ground-state properties and the DDMRG method~\cite{eric} 
to calculate dynamical properties. In order to carry out our calculations, we 
consider $N+2$ electrons in a system consisting of $N$ noninteracting bath sites 
($N$ even) and two impurity sites. The electron density is 
$\left\langle n_\uparrow \right\rangle = \left\langle n_\downarrow \right\rangle = N/2+1$. 

The (D)DMRG calculations can be performed on finite lattices only, i.e., we must 
discretize the host band and carry out (D)DMRG calculations for finite number $N$ of host band eigenstates 
corresponding to energies $\varepsilon_k$ ($k = 1,...,N$), and then 
extrapolate the results to a continuous host band 
($N \to \infty$) if needed. Choosing a discretization of the 
host band, i.e., selecting the $N$ band state energies 
$\varepsilon_k$, should be done appropriately depending on what is
to be obtained.

\begin{figure}
\begin{center}
\includegraphics[width=9.5cm,clip]{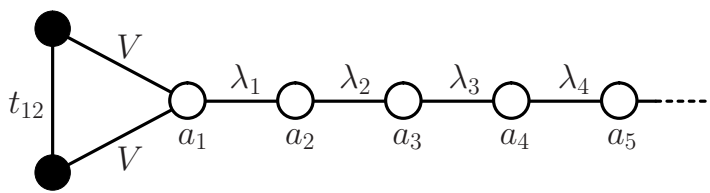}
\caption{One-dimensional lattice configuration for applying the DMRG
  algorithm to the two-impurity problem. 
  The solid circles denote the impurity sites and the 
  open circles denote the host band.}
\label{fig_model}
\end{center}
\end{figure}

The Hamiltonian (\ref{hamiltonian}) is, however, somewhat unsuited for a DMRG 
treatment because it includes hopping terms that are long-range. 
For example, 
the system size is limited to $N \le 60$ for typical calculations
even when several thousand density-matrix eigenstates are kept
in the DMRG procedure, the maximum possible on current
workstations. 
Therefore, we first transform the Hamiltonian (\ref{hamiltonian}) 
into a linear chain with nearest-neighbor hopping only,
\begin{eqnarray}
\nonumber
\hat{H} &=& 
V \sum_{i\sigma} \left( \hat{c}_{0\sigma}^{\dag}\hat{d}^{\phantom{\dagger}}_{i\sigma} +
\hat{d}_{\sigma}^{\dag}\hat{c}^{\phantom{\dagger}}_{0\sigma}\right) 
\nonumber
+ \sum_{j\sigma} a_j \hat{c}_{j\sigma}^{\dag} \hat{c}^{\phantom{\dagger}}_{j\sigma}
+ \sum_{j\sigma} \lambda_{j}
\left( \hat{c}_{j\sigma}^{\dag}\hat{c}^{\phantom{\dagger}}_{j+1\sigma} +
\hat{c}_{j+1\sigma}^{\dag}\hat{c}^{\phantom{\dagger}}_{j\sigma}\right) \\
&+& U \sum_{i=1,2} \left(\hat{n}^d_{i\uparrow} - \frac{1}{2}\right)
\left(\hat{n}^d_{i\downarrow} - \frac{1}{2}\right) 
+ t_{12} \sum_{\sigma}
(\hat{d}_{1\sigma}^{\dag}\hat{d}^{\phantom{\dagger}}_{2\sigma} 
+ h.c.)\, . 
\label{hamiltonian2}
\end{eqnarray}
The new fermion operators $\hat{c}^{\phantom{\dagger}}_{j\sigma}$ correspond to electronic 
states in the host band and are related to the original representation 
by a canonical transformation
\begin{eqnarray}
\hat{c}_{j\sigma}^{\phantom{\dagger}} = \sum_k M_{jk} \hat{f}_{k\sigma}^{\phantom{\dagger}}.
\label{canonical}
\end{eqnarray}
The orthogonal matrix $M_{jk}$, the diagonal terms $a_j$, and the
nearest-neighbor hopping terms $\lambda_j$ are calculated using the
Lanczos algorithm for tridiagonalizing a symmetric matrix starting from 
the initial vector ${M_{1,k} = V_k/V}$ with $V^2 = \sum_k V_k^2$. For a 
hybridization function (\ref{hybridization}) symmetric about 
$\omega=0$, the diagonal terms $a_j$ vanish. The Hamiltonian 
(\ref{hamiltonian2}) describes two impurities coupled to one end of a 
one-dimensional chain representing the host band states 
(see Figure~\ref{fig_model}). This transformation enables us to handle a system 
with up to $N \sim {\cal O}(200)$ bath states using the 
(D)DMRG method. 

Furthermore, for efficient treatment, we introduce even- ($p=e$) and odd- 
($p=o$) parity impurity orbitals 
$d_{p\sigma}=(d_{1\sigma} \pm d_{2\sigma})/\sqrt{2}$ as in Ref.19. With 
${\bf R}_1-{\bf R}_2={\bf 0}$, the Hamiltonian (\ref{hamiltonian}) is 
transformed to
\begin{eqnarray}
\hat{H} &=& \hat{H}_0 + \hat{H}_U \\
\nonumber
\hat{H}_0 &=& V \sum_\sigma (\hat{d}_{e\sigma}^{\dag} \hat{c}^{\phantom{\dagger}}_{0\sigma} + h.c.) 
+ \sum_{j\sigma} \lambda_{j}(\hat{c}_{j\sigma}^{\dag} 
\hat{c}^{\phantom{\dagger}}_{j+1\sigma} + h.c.) \\
&+& t_{12} \sum_\sigma (\hat{d}_{e\sigma}^{\dag} \hat{d}^{\phantom{\dagger}}_{e\sigma} 
- \hat{d}_{o\sigma}^{\dag} \hat{d}^{\phantom{\dagger}}_{o\sigma}) \\
\label{ham3_t}
\nonumber
\hat{H}_U &=& \frac{U}{2} 
[(\hat{n}_{e\uparrow}+\hat{n}_{o\uparrow}-1)
(\hat{n}_{e\downarrow}+\hat{n}_{o\downarrow}-1) \\
&+& (\hat{d}_{e\uparrow}^{\dag} \hat{d}_{o\uparrow} 
+ \hat{d}_{o\uparrow}^{\dag} \hat{d}^{\phantom{\dagger}}_{e\uparrow})
(\hat{d}_{e\downarrow}^{\dag} \hat{d}^{\phantom{\dagger}}_{o\downarrow} 
+ \hat{d}_{o\downarrow}^{\dag}
\hat{d}^{\phantom{\dagger}}_{e\downarrow})] 
\label{ham3_U} \, ,
\end{eqnarray}
where $\hat{n}_{e(o)\sigma}=\hat{d}_{e(o)\sigma}^{\dag}\hat{d}^{\phantom{\dagger}}_{e(o)\sigma}\,$. 
Note that only the even-parity orbital hybridizes with the noninteracting 
bath states directly. When $U=0$, the even- and odd-parity orbitals are 
completely separate. 
In our model, they will be mixed only via the Coulomb interaction.

\section{Results}

\subsection{Static properties}

We begin our discussion with some static properties of the system,
namely, the electron
density at the impurities and the spin-spin correlation function
between the impurities, calculated with the standard DMRG method. 
Here we apply a logarithmic 
discretization scheme of the host band, $\varepsilon_k=(W/2)\Lambda^{-k}$ 
(with $\Lambda>1$ and $k=1,2,\cdots, N/2$), as usually used in Wilson's 
renormalization group method~\cite{wilson}, because we are interested
in the case of large host bandwidth
as well as in having dense bath states around the chemical 
potential for quantitative accuracy. Typically, we use $N=38$ bath states with 
$\Lambda=1.5$ and $W=100\pi\Delta_0$ and keep $m=2000$ density-matrix 
eigenstates in the DMRG procedure. In some cases, systems with up to $N=58$ 
and $W=1000\pi\Delta_0$ are used to extrapolate the results to $W \to \infty$.

\subsubsection{Electron density at the impurity}

\begin{figure}[t]
\begin{center}
\vspace{5pt}
\includegraphics[width=8.0cm,clip]{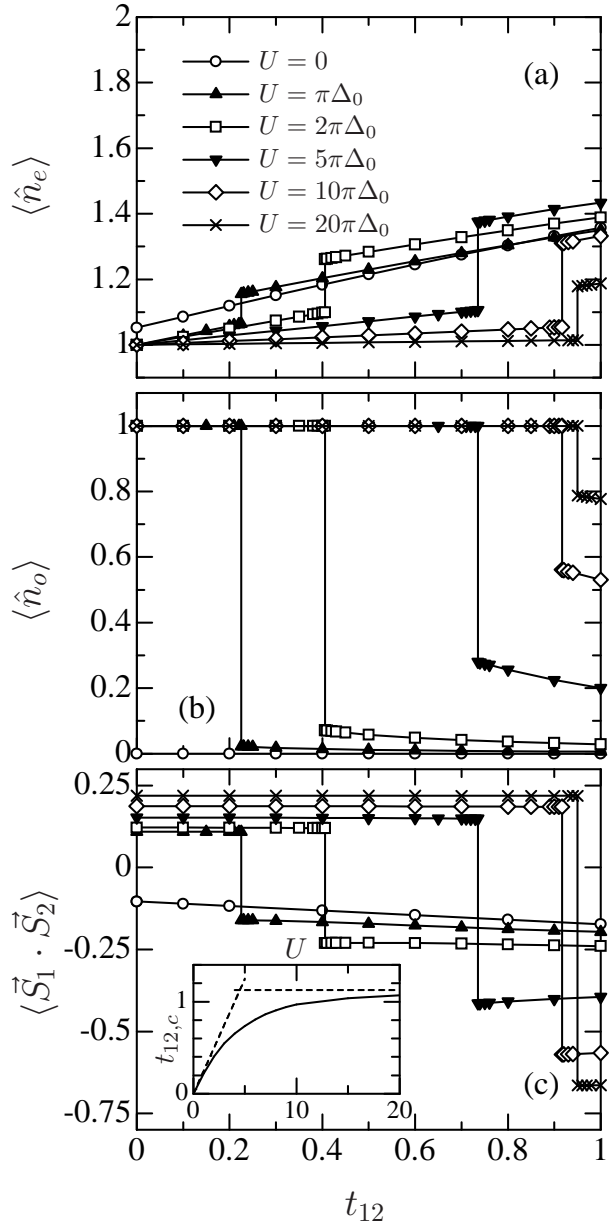}
\vspace{-5pt}
\caption{Local density on the even- (a) and odd- (b) parity orbitals of the 
coupled impurities as a function of $t_{12}$ for several $U$ values. (c) 
Spin-spin correlation function as a function of $t_{12}$ and $U$. Inset: the 
critical values of the transition, $t_{12,c}$, as a function of $U$. The dashed 
lines correspond to $t_{12,c}=U/4$ and $t_{12,c}=1.13$ (see text).}
\label{fig2}
\end{center}
\end{figure}

The average electron densities $\left\langle \hat{n}_e \right\rangle$ and 
$\left\langle \hat{n}_o \right\rangle$ for even- and odd-parity orbitals of 
coupled impurities, respectively, are displayed in Fig.~\ref{fig2}(a) and (b)
for different values of $U$ as function of $t_{12}$.
Since the two-impurity sites are equivalent, we have
$\left\langle \hat{n}_1 \right\rangle = \left\langle \hat{n}_2 \right\rangle
=(\left\langle \hat{n}_e \right\rangle + \left\langle \hat{n}_o \right\rangle)/2$. 
Note that $\left\langle \hat{n}_1 \right\rangle (\left\langle \hat{n}_2 \right\rangle)$ 
can take values between 0 and 2 due to the charge degrees of freedom at 
the impurity sites, in contrast to the two-impurity Kondo model for which 
$\left\langle \hat{n}_1 \right\rangle = \left\langle \hat{n}_2
\right\rangle =1$ always.

When $t_{12}=0$, 
$\left\langle \hat{n}_e \right\rangle = \left\langle \hat{n}_o \right\rangle = 1$ 
for all interaction strengths due to particle-hole symmetry.
The local densities are drastically affected by 
$t_{12}$. We find a discontinuous transition at a critical value 
$t_{12}=t_{12,c}$ which is dependent on $U$. The critical value $t_{12,c}$ is 
zero at $U=0$ and becomes 
larger with increasing $U$. In the large $U$ limit, it saturates at 
$t_{12,c} \sim \pi\Delta_0$. 
We show the critical value $t_{12,c}$ as a 
function of $U$ for $W \to \infty$  in the inset of Fig.~\ref{fig2}(c). 

At $U=0$, $\left\langle \hat{n}_o \right\rangle$ drops to zero for
infinitesimally small 
$t_{12}(<0)$. In this limit, there is no hybridization with the conduction 
band and no coupling to the even-parity level, i.e., the odd-parity orbital forms a completely local state at the Fermi level 
$\varepsilon_{\rm F}$ iff $t_{12}=0$. An infinitesimally small $t_{12}$ shifts the 
state above $\varepsilon_{\rm F}$, i.e., $t_{12,c}=0^+$. For small but finite 
$U$, the situation is similar and the transition is caused by the competition 
between $t_{12}$ and $U$. At $t_{12}=0$, the odd-parity orbital is
split by the Coulomb interaction [see Eq.(\ref{ham3_U})] into 
two states with energy difference $\sim U/2$, the lower and the upper 
Hubbard band (we call them the LHB and the UHB, respectively) located at 
$\omega \approx \pm U/4$. A finite $t_{12}$ has the effect of 
shifting the LHB to higher energies by 
$|t_{12}|$. The odd-parity orbital is occupied by one electron 
if the LHB is below $\varepsilon_F$ and almost vacant if the LHB is above 
$\varepsilon_F$. This leads to $t_{12,c} \sim U/4$ and a large discontinuity 
($\sim 1$) of $\left\langle \hat{n}_o \right\rangle$ at $t_{12,c}$. 
This anticipated behavior is in fact observed in our DMRG results. As $U$ increases, the 
odd-parity orbital begins to hybridize indirectly with the host band via the 
interaction term (\ref{ham3_U}). The discontinuity, therefore, becomes 
smaller and goes to zero as $U \to \infty$. 
Moreover, we find that $t_{12,c}$ has
almost no $U$-dependence in the large-$U$ regime. 
This implies that 
the physics behind the transition for large $U$ is different from the
competition between $t_{12}$ and $U$ at small $U$. Before we discuss a
possible mechanism in the next paragraph, let us briefly discuss the behavior
of the even-parity orbital. For all $U$ values, 
$\left\langle \hat{n}_e \right\rangle$ increases slowly as a function of 
$t_{12}$ because the even-parity orbital hybridizes 
directly with the host band, thus experiencing a strong broadening. 
Hence, the discontinuity 
of $\left\langle \hat{n}_e \right\rangle$ at $t_{12}=t_{12,c}$ is
smaller than  
that of $\left\langle \hat{n}_o \right\rangle$. 
We can also see the largest 
discontinuity, which means that the charge fluctuation on the
even-parity orbital is largest, for intermediate $U$ values 
($U \sim 5\pi \Delta_0$). We also note  
that $\left\langle \hat{n}_o \right\rangle=1$ independent of $U$ 
below $t_{12,c}$, and $\left\langle \hat{n}_o \right\rangle \to 0$ and 
$\left\langle \hat{n}_e \right\rangle \to 2$ as $t_{12} \to \infty$
for finite $U$.

\subsubsection{Spin-spin correlation between impurities}

We now investigate the spin-spin correlation between the two impurities, 
$\left\langle \vec{S}_1 \cdot \vec{S}_2 \right\rangle$. The results 
are shown in Fig.~\ref{fig2}(c) as a function of $t_{12}$ for
different values of $U$. 
Before we discuss the results, let us first identify the different  
types of magnetic interactions present in our model.
First, we have the $c\,$-$f$ exchange interaction $J_{cf}$, 
which is the antiferromagnetic interaction between an electron on the impurity 
sites and conduction electrons. For a single impurity, $J_{cf}$ leads
to the Kondo effect and a {\em local} spin singlet as the ground
state. Within the standard Schrieffer-Wolff mapping, the value for
$J_{cf}$ is given by $J_{cf}=8\Delta_0/U_{\rm eff}$, 
where $U_{\rm eff}$ is the effective Coulomb interaction and 
the impurities are both occupied by one electron. The $c\,$-$f$ exchange is
effective 
only when an electron is localized on the impurity so that $U_{\rm eff}=U/2$ 
and $J_{cf}=16\Delta_0/U$. 
Second, the conduction electrons mediate the RKKY 
interaction $J_{\rm RKKY}$. For our particular setup, we obtain a ferromagnetic interaction 
between two electrons on the impurity sites. As usual, $J_{\rm RKKY}$
is obtained as second-order process, i.e., $J_{\rm RKKY} \sim J_{cf}^2
\sim {\cal O}(1/U^2)$. Third, the model exhibits a
direct exchange interaction $J_{\rm ex}$ due to the coupling $t_{12}$, which is 
an antiferromagnetic interaction and is given by the standard expression
$J_{\rm ex} = 4t_{12}^2/U$.

We expect that for $t_{12} < t_{12,c}$ ferromagnetic correlations due to the RKKY
interaction are dominant, i.e.,  
$\left\langle \vec{S}_1 \cdot \vec{S}_2 \right\rangle>0$. 
On the other hand, when $t_{12} > t_{12,c}$, 
antiferromagnetic correlations due to the exchange interaction are
stronger, i.e.,  
$\left\langle \vec{S}_1 \cdot \vec{S}_2 \right\rangle<0$. 
Such a transition from ferromagnetic to antiferromagnetic correlations
at $t_{12,c}$ is in fact found at  all interaction strengths, as can be seen
in Fig.~\ref{fig2}(c). 
The absolute value of  
$\left\langle \vec{S}_1 \cdot \vec{S}_2 \right\rangle$ increases with increasing 
$U$ and reaches the maximum possible value as $U \to \infty$, which
means that one electron is localized on each impurity in the
$U\to\infty$ limit.

Let us now consider the $U$-dependence of the critical value $t_{12,c}$. The 
transition is driven by the charge degrees of freedom in the weak-coupling 
regime. However, the spin degrees of freedom play an essential role in the 
strong-coupling regime. Taking into account that the spin-spin correlations between the 
impurity sites change from ferromagnetic to antiferromagnetic at 
$t_{12,c}$, we may expect that the competition between the RKKY and
direct exchange  
interactions is the origin of the transition. If we take $J_{\rm ex}=J_{\rm
RKKY}$ as the criterion for the occurrence of the transition, we obtain 
$t_{12,c} \sim {\cal O}(1/\sqrt{U})$.
Thus, $t_{12,c}$ would go to zero as $U \to \infty$, which is obviously
inconsistent with the DMRG results which show almost 
constant $t_{12,c}$ as a function of $U$. 

Up to now, we have not taken into account $J_{cf}$, which leads to a
competition between the formation of local Kondo and non-local
singlets, offering a quite different mechanism for the
transition. In this case, the criterion to obtain 
$t_{12,c}$ is $J_{cf}=J_{\rm ex}$, which gives 
$t_{12,c} = \sqrt{4/\pi} \sim 1.13$. This result is indeed consistent
with our findings. Thus, the transition in the strong coupling regime
can be interpreted as competition between local singlet formation due
to the Kondo effect and non-local singlet formation due to the direct
exchange introduced by $t_{12}$. Note, however, that, in contrast to
the general folklore,  the boundary is {\em not} set by 
$T_K(J_{cf})=J_{\rm ex}$, but by the direct comparison of the bare
energy scales.
We also find that the total spin is $S=1$ for $t_{12} < t_{12,c}$ and
$S=0$ for $t_{12} > t_{12,c}\,$.

\subsection{Dynamical properties}

In this section, we study the spectral density for impurities in the TIAM. 
The impurity one-particle Green function for even- and odd-parity orbitals 
can be written as
\begin{eqnarray}
G_{p\sigma}(\omega) &=& \left \langle \hat{d}^\dag_{p\sigma} 
\frac{1}{\hat{H}-E_0+\omega-i\eta} \hat{d}^{\phantom{\dagger}}_{p\sigma} \right \rangle 
+ \left \langle \hat{d}^{\phantom{\dagger}}_{p\sigma} 
\frac{1}{E_0-\hat{H}+\omega+i\eta} \hat{d}^\dag_{p\sigma} \right \rangle
\label{green}
\end{eqnarray}
($\eta \rightarrow 0^+$), where $E_0$ is the ground-state energy
and $\langle \dots \rangle$ represents a ground-state expectation value.
The impurity spectral density for each parity is then obtained as
\begin{equation}
D_{p\sigma}(\omega) = -\frac{1}{\pi} \, {\rm sgn}(\omega)
{\rm Im} \, G_{p\sigma}(\omega)
= A_{p\sigma}(\omega) + B_{p\sigma}(\omega)
\label{DOS}
\end{equation}
with 
\begin{eqnarray}
\nonumber
A_{p\sigma}(\omega \leq 0) & = & \lim_{\eta \rightarrow 0}
\left \langle \hat{d}^\dag_{p\sigma}
\frac{\eta}{\pi[(\hat{H}-E_0+\omega)^2+\eta^2]} \hat{d}^{\phantom{\dagger}}_{p\sigma} \right 
\rangle \label{ADOS}\\
\nonumber
B_{p\sigma}(\omega \geq 0) & = & \lim_{\eta \rightarrow 0}
\left \langle \hat{d}^{\phantom{\dagger}}_{p\sigma} 
\frac{\eta}{\pi[(\hat{H}-E_0-\omega)^2+\eta^2]} 
\hat{d}^\dag_{p\sigma} \right \rangle 
\label{BDOS}
\end{eqnarray}
and $A_{p\sigma}(\omega \geq 0) = B_{p\sigma}(\omega \leq 0) =0$.
The spectral density fulfills the sum rule 
\begin{equation}
\int_{-\infty}^{\infty} D_{p\sigma}(\omega) {\rm d} \omega  =  1 .
\label{sumrule}
\end{equation}
Note that the spectral densities for both impurities are the same, i.e., 
$D_{1\sigma}(\omega)=D_{2\sigma}(\omega)$, and are equal to 
$[D_{o\sigma}(\omega)+D_{e\sigma}(\omega)]/2$. 

The standard DMRG algorithm~\cite{steve,dmrgbook} can be used to calculate 
the ground-state properties as shown in the last section. In particular, 
the ground-state wave function $|\Psi_0\rangle$ and the ground-state energy 
$E_0$ can readily be obtained. To compute dynamic properties such as the 
impurity Green's function~(\ref{green}) we use the DDMRG~\cite{eric}. This 
approach is based on a variational principle. One can easily show that for 
$\eta > 0$ and fixed frequency $\omega$, the minimum of the functional
\begin{eqnarray}
\nonumber
W(\Psi) &=&
\left \langle \Psi \left | \left(E_0+\omega-\hat{H}\right)^2 +\eta^2
\right | \Psi 
\right \rangle \\
&+& \eta \left \langle \Psi_0 \left | \hat{d}^{\phantom{\dagger}}_{p\sigma} \right | 
 \Psi \right \rangle
+ \eta  \left \langle \Psi \left | \hat{d}_{p\sigma}^\dagger 
\right | \Psi_0  \right \rangle
\label{functional}
\end{eqnarray}
with respect to all quantum states $|\Psi\rangle$ is
\begin{equation}
W(\Psi_{\rm min}) = 
\left \langle \Psi_0 \left | \hat{d}^{\phantom{\dagger}}_{p\sigma} 
\frac{-\eta^2}{\left(E_0+\omega-\hat{H}\right)^2 +\eta^2}
\hat{d}_{p\sigma}^\dagger \right | \Psi_0 \right \rangle .
\end{equation} 
The functional minimum is related  to the convolution of the spectral 
density~(\ref{BDOS}) with a Lorentz distribution of width $\eta$ by
\begin{equation}
W(\Psi_{\rm min}) = -\pi \eta B_{p\sigma}^{\eta}(\omega).  
\end{equation}
A similar result is obtained for the spectral density~(\ref{ADOS}) 
if one substitutes $\hat{d}_{p\sigma}^{\phantom{\dagger}}$
for $\hat{d}^\dagger_{p\sigma}$, $-\omega$ for $\omega$ and 
$A_{p\sigma}^{\eta}(\omega)$ for $B_{p\sigma}^{\eta}(\omega)$
in the above equations.

The DDMRG method consists essentially of  
minimizing the functional~(\ref{functional}) numerically using
the standard DMRG algorithm.
Thus the DDMRG provides the spectral densities
$A^{\eta}_{p\sigma}(\omega)$ and $B^{\eta}_{p\sigma}(\omega)$
for a finite broadening $\eta$.
The full spectral density~(\ref{DOS}) convolved with the Lorentz
distribution 
\begin{equation}
D^{\eta}_{p\sigma}(\omega) =  \int_{-\infty}^{\infty} {\rm d}\omega^\prime
D_{p\sigma}(\omega') \frac{\eta}{\pi[(\omega-\omega')^2+\eta^2]}
\label{convolution}
\end{equation}
is given by the sum of  $A^{\eta}_{p\sigma}(\omega)$ and 
$B^{\eta}_{p\sigma}(\omega)$.
The real part of the Green's function can be calculated 
with no additional computational cost but is generally less accurate.
The necessary broadening
of spectral functions in DDMRG calculations
is actually very useful for studying continuous spectra
or for doing a finite-size scaling analysis~\cite{eric}.

What we really would like to obtain is the spectral density in the $\eta=0$ limit. 
This can be done by carrying out a 
deconvolution of the DDMRG data~\cite{nishim}. In theory, a deconvolution 
amounts to solving (\ref{convolution}) for $D_{p\sigma}(\omega)$ using
the DDMRG
data on the left-hand side. We also know that the broadened spectral density 
of the impurity system on an infinite lattice ($N \to \infty$) is usually almost 
identical to the spectral density of the discretized impurity system 
($N < \infty$) if $\eta \ge \Delta\varepsilon$. Therefore, one can make 
the approximation that the DDMRG data for $D^{\eta}_{p\sigma}(\omega)$
describes the broadened spectral density for $N \to \infty$ and can
then solve 
(\ref{convolution}) approximately under the condition that $D_{p\sigma}(\omega)$ 
is the exact spectral density of the TIAM. For instance, one can require 
that $D_{p\sigma}(\omega)$ is a continuous and relatively smooth function. 
To obtain quantitatively accurate spectra after deconvolution, we need to 
take $\eta$ smaller than the width of the spectra in $\eta=0$. 

\subsubsection{noninteracting case}

\begin{figure}[t]
\begin{center}
\vspace{5pt}
\includegraphics[width=8.0cm,clip]{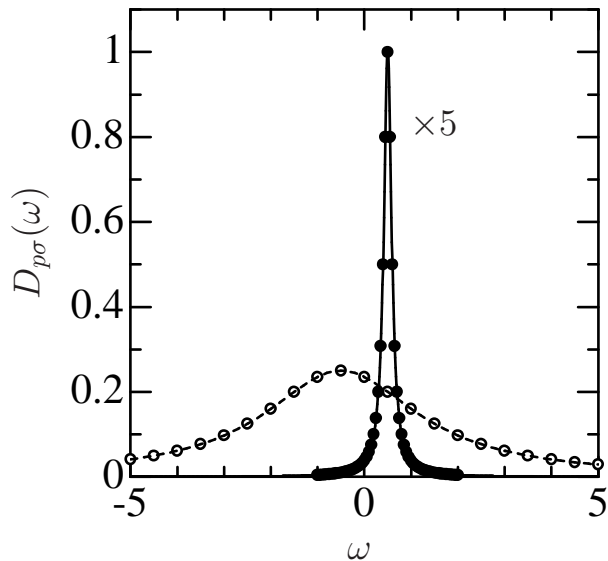}
\vspace{-5pt}
\end{center}
\caption{Spectral density at $U=0$ and $W=20\pi\Delta_0$. The open circles 
denote $D_{e\sigma}(\omega)$ calculated with a constant host band 
discretization for $N=59$, $\Delta \varepsilon \approx 0.34\pi \Delta_0$, and 
$\eta=0.5\pi\Delta_0$, then deconvolved. Solid circles denote 
$D_{e\sigma}(\omega)$ calculated with variable discretization 
$\Delta \varepsilon \approx 0.067\pi \Delta_0$ and $\eta=0.1\pi\Delta_0$ 
around the peak. Solid and dashed lines are the exact solutions (\ref{oddu0}) 
with broadening $\eta=0.1\pi\Delta_0$ and (\ref{evenu0}) without broadening, 
respectively.}
\label{fig_u0}
\end{figure}

The spectral density at $U=0$ can be calculated exactly and
provides a good test to demonstrate the accuracy of our method. 
The exact spectral density for the odd-parity orbital is a
a $\delta$-function at $\omega=-t_{12}$
\begin{equation}
D_o(\omega)=\delta(\omega+t_{12}),
\label{oddu0}
\end{equation}
while for the even-parity orbital we obtain a Lorentzian of width $2\Delta_0$ centered at $\omega=t_{12}$
\begin{equation}
D_e(\omega)=\frac{2\Delta_0}{\pi[(\omega-t_{12})^2+(2\Delta_0)^2]}\;.
\label{evenu0}
\end{equation}
In Fig.~\ref{fig_u0}, we show the spectral density 
calculated with the DDMRG for $U=0$. On the scale of Fig.~\ref{fig_u0}, 
there is no visible difference between our numerical results and the
exact results. The deconvolution technique is not useful for obtaining a 
divergent function such as a $\delta$-function, so we introduce a finite 
broadening $\eta=0.1\pi\Delta_0$ into Eq.(\ref{oddu0}) and compare it to our `bare' 
DDMRG spectra convolved with same $\eta$. Note that the local Coulomb 
interaction $U$ is always treated numerically exactly in the
density-matrix renormalization and 
thus does not affect the accuracy of the method directly. 

\subsubsection{weak-coupling regime}

\begin{figure}[t]
\begin{center}
\vspace{5pt}
\includegraphics[width=7.5cm,clip]{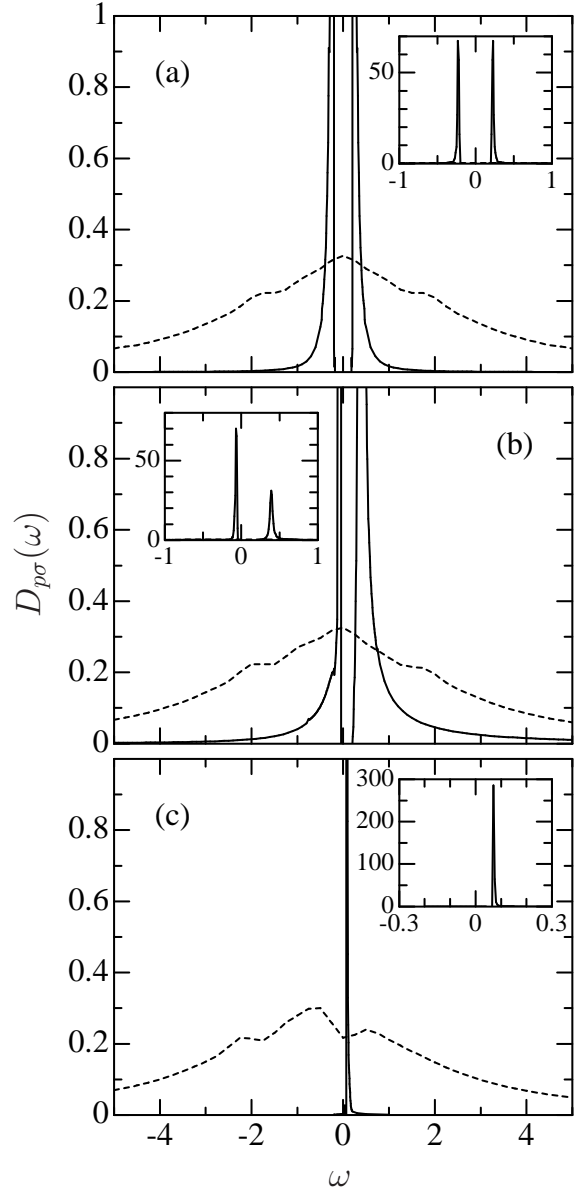}
\vspace{-5pt}
\end{center}
\caption{Spectral density of the coupled impurities for the odd-parity orbital 
$D_{o\sigma}(\omega)$ and for the even-parity orbital $D_{e\sigma}(\omega)$ 
with (a) $t_{12}=0$, (b) $t_{12}=0.15\pi\Delta_0$, and (c) 
$t_{12}=0.25\pi\Delta_0$ at 
$U=\pi\Delta_0$ and $W=20\pi\Delta_0$. Dashed lines denote 
$D_{e\sigma}(\omega)$ calculated with a constant host band 
discretization for $N=58$, $\Delta \varepsilon \approx 0.34\pi \Delta_0$, and 
$\eta=0.5\pi\Delta_0$, then deconvolved. Solid lines denote 
$D_{o\sigma}(\omega)$ calculated with a variable discretization for $N=118$ 
($N=70$), $0.01 \pi \le \Delta \varepsilon/\Delta_0 \le 0.45\pi$ 
($0.0033 \pi \le \Delta \varepsilon/\Delta_0 \le 1.98\pi$)
and a constant broadening $\eta=0.02\pi\Delta_0$ ($\eta=0.005\pi\Delta_0$) 
for $t_{12}=0$ and $0.15\pi\Delta_0$ ($0.25\pi\Delta_0$), then deconvolved. 
Insets: expanded view around the Fermi level $\omega=0$.}
\label{fig_u1}
\end{figure}

In the weak-coupling regime ($U \ll 4\pi\Delta_0$), the physical
properties are still 
similar to those of the noninteracting case. The spectral density of the impurities 
calculated with the DDMRG for $t_{12}/\Delta_0=0$, $0.15\pi$, and $0.25\pi$ at 
$U=\pi\Delta_0$ is shown in Fig.~\ref{fig_u1}. The critical coupling $t_{12,c}$ is 
$0.205\pi\Delta_0$. Let us first look at the spectrum for the even-parity orbital, 
$D_{e\sigma}(\omega)$. At $t_{12}=0$ 
[see Fig.~\ref{fig_u1}(a)], $D_{e\sigma}(\omega)$ is basically a 
Lorentzian of width $\sim 2 \Delta_0$ centered at $\omega=0$, but there 
appear small shoulders around $\omega \sim \pm0.8\pi\Delta_0 (\sim U)$ and 
$\omega \sim \pm1.8\pi\Delta_0 (\sim 2U)$ due to the Coulomb interaction 
(\ref{ham3_U}). When $t_{12}$ increases, the central peak is 
shifted towards lower energies by $|t_{12}|$ while maintaining its shape 
[see Fig.~\ref{fig_u1}(b) and (c)]. This is consistent with the gradual increase of 
the local density on the even-parity orbital $\left\langle \hat{n}_e \right\rangle$ 
as a function of $t_{12}$ and also with the small discontinuity at
$t_{12,c}$, as shown in Fig.~\ref{fig2}(a). 
We next turn to the spectrum for 
the odd-parity orbital, $D_{o\sigma}(\omega)$. As long as $U=0$, this orbital has no direct hybridization 
with the host band, so that $D_{o\sigma}(\omega)$ consists of a
localized state, i.e.,
a single $\delta$-function peak. When a small $U (\le \pi\Delta_0)$ 
is introduced at $t_{12}=0$, this peak splits into two peaks located around 
$\sim \pm U/4$ due to the effective repulsion on the odd-parity orbital 
$U_{\rm eff}=U/2$ [see Eq.(\ref{ham3_U})]. They correspond to the LHB and 
the UHB. In other words, the odd-parity orbital is half-filled and a
Mott-Hubbard gap opens. The peaks are still very sharp but are no longer exact 
$\delta$-functions, as we can see in Fig.~\ref{fig_u1}(a), because the odd-parity 
orbital couples `indirectly' to the host band via the even-parity orbital. As 
$t_{12}$ increases, the two peaks are shifted towards higher energies by 
$|t_{12}|$, but their separation remains $\sim U/2$. The LHB and UHB become 
sharper (broader) while retaining their respective weights. 
In addition, no spectral weight is transferred to the gap.
When the LHB reaches the Fermi level 
$\varepsilon_F$ ($\omega=0$) at $t_{12}=t_{12,c}(\sim U/4)$, the transition 
occurs. When $t_{12} > t_{12,c}$ [see Fig.~\ref{fig_u1}(c)], 
only one peak is present, and it is above $\omega=0$ and is very sharp. 
Actually, 
there must be some spectral weight below $\omega=0$ because
$\left\langle \hat{n}_o \right\rangle$ is not exactly zero [see Fig.~\ref{fig2}(b)]. 
We therefore find 
that the odd-parity orbital behaves as a nearly localized state for
all $t_{12}$  
and that the charge degrees of freedom in the odd-parity orbital play a 
crucial role for the transition.

\subsubsection{strong-coupling regime}

\begin{figure}[t]
\begin{center}
\vspace{5pt}
\includegraphics[width=7.5cm,clip]{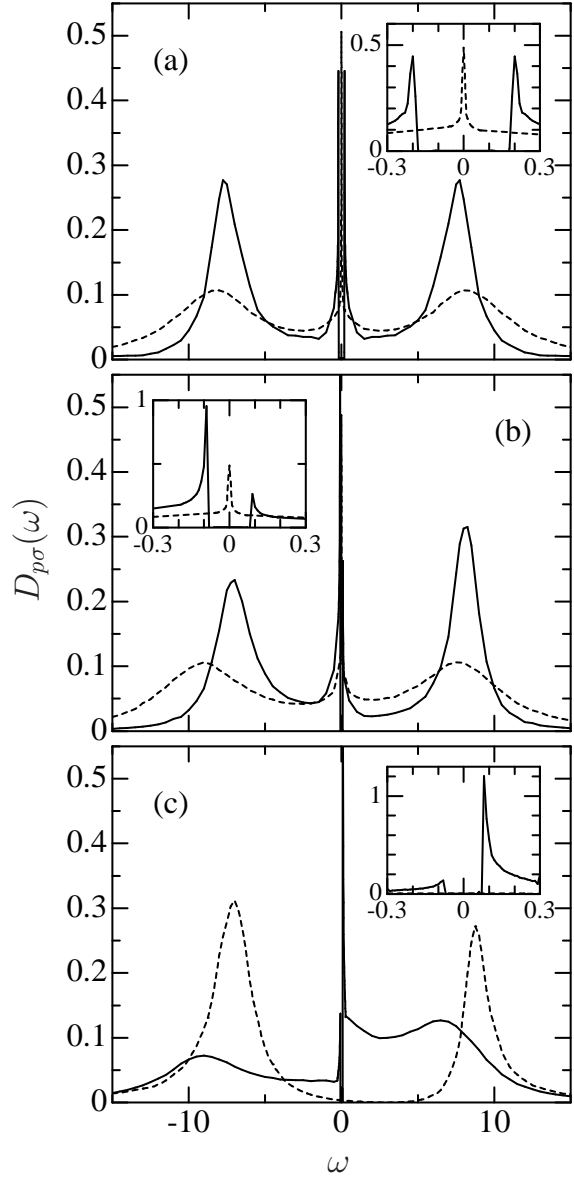}
\vspace{-5pt}
\caption{Spectral density of the coupled impurities for the odd-parity orbital 
$D_{o\sigma}(\omega)$ and for the even-parity orbital $D_{e\sigma}(\omega)$ 
with (a) $t_{12}=0$, (b) $t_{12}=0.6\pi\Delta_0$, and (c) $t_{12}=\pi\Delta_0$ at 
$U=15\pi\Delta_0$ and $W=40\pi\Delta_0$. Dashed lines denote 
$D_{e\sigma}(\omega)$ calculated with a variable discretization for $N=94$, 
$\Delta \varepsilon \approx 0.0013\pi \Delta_0$, and $\eta=0.002\pi\Delta_0$, 
then deconvolved. Solid lines denote $D_{o\sigma}(\omega)$ calculated with a 
variable discretization for $N=94$, $0.0068 \pi \le \Delta \varepsilon/\Delta_0 \le 1.98\pi$ 
and a constant broadening $\eta=0.01\pi\Delta_0$, then deconvolved. 
Insets: expanded view around the Fermi level $\omega=0$.}
\label{fig_u15}
\end{center}
\end{figure}

We now consider the spectral density in the strong-coupling regime 
($U \gg 4\pi\Delta_0$), which corresponds to the so-called Kondo
regime in the SIAM. 
For the TIAM, however, the situation is more complex due to competing 
interactions, i.e., the RKKY interaction, the Kondo (or $c\,$-$f$ exchange) 
effect, and the exchange interaction between impurity sites. 
In Fig.~\ref{fig_u15}, we show the spectral density calculated with the 
DDMRG for $t_{12}=0$, $0.6\pi\Delta_0$, and $\pi\Delta_0$ at $U=15\pi\Delta_0$. 
The critical inter-impurity hopping $t_{12,c}\sim 0.8\pi\Delta_0$ is
slightly smaller than that obtained in Sec.\ II because the
calculations were done for a finite 
host bandwidth here. Let us first look at the even-parity 
spectral density $D_{e\sigma}(\omega)$. Below $t_{12,c}$, we can see a sharp 
peak at $\omega=0$ in $D_{e\sigma}(\omega)$, which satisfies the Friedel sum rule 
$D_{e\sigma}(\omega=0)=1/(2\pi\Delta_0)$. This means that the conduction 
electrons form a spin-singlet (Kondo) state with electrons in the 
even-parity orbital. 
The width of this peak at $\omega=0$ becomes smaller exponentially 
with increasing $U$. We can thus state that the properties of 
$D_{e\sigma}(\omega)$ below $t_{12,c}$ are similar to that of the SIAM which 
is characterized by the Abrikosov-Suhl resonance at $\omega=0$ and the Hubbard 
satellites around $\omega \approx \pm U/2$. Moreover, we notice that the 
physics of the TIAM is quite different from that of the two-impurity Kondo 
model, where no Kondo effect is observed~\cite{vojta}, at least for the case of equivalent 
impurities. Note also that in the 
TIAM, states with only one electron on the two impurity sites can still
have a large weight ($\sim 3\%$ for $U=15\pi\Delta_0$) in the eigenvector of 
the ground-state even in the Kondo regime, in contrast to the 
two-impurity Kondo model. 

When we increase $t_{12}$, the 
shape of $D_{e\sigma}(\omega)$ is hardly changed and only a weak
transfer of spectral weight from above $\varepsilon_F$ to below 
$\varepsilon_F$ occurs, consistent with the behavior of $\left\langle \hat{n}_e \right\rangle$. 
As long as $t_{12}$ is smaller than $t_{12,c}$, the quasi-particle
peak stays pinned at $\omega=0$ and maintains its height
($1/(\pi\Delta_0)$), while the Hubbard 
satellites stay located at $|\omega| < U/2$ with width $> 2\Delta_0$. 
Above $t_{12,c}$, on the 
other hand, the Kondo peak vanishes. The 
local spin is now screened due to the formation of a non-local 
singlet between the impurities due to the dominant exchange
interaction, i.e., the scattering channels leading to the
Abrikosov-Suhl resonance are not active any more.
Simultaneously, the `effective' hybridization between the host band and the 
even-parity orbital becomes weaker so that the width of Hubbard
satellites becomes narrower. Our numerical results indicate that the
Hubbard satellites turn into Lorenzians with width $2\Delta_0$ and weight 
$1/2$ located at $|\omega| = U/2$ in the limit of $t_{12} \to \infty$ 
and $U \to \infty$.

Next, let us turn 
to the odd-parity spectral density $D_{o\sigma}(\omega)$. At $t_{12}=0$, we 
find four peaks. The two prominent peaks located at 
$\omega \approx \pm 7.5 ( = \pm \pi\Delta_0U/2)$ can be identified
with the Hubbard bands of the electrons localized in this
orbital. Since no direct hybridization to the band states exists,
these peaks are sharper than the corresponding ones in
$D_{e\sigma}(\omega)$, i.e., the broadening is introduced indirectly via the
Coulomb interaction.
Rather more interesting is the appearance of a structure at the Fermi
energy consisting of two narrow peaks separated by a gap $\Delta \sim
0.36\approx J_{cf}=16/(\pi U) \approx 0.34$ for $t_{12}=0$. As $t_{12}$
increases, this feature and, in particular, the gap prevails, although
its size decreases (see insets to Fig.~\ref{fig_u15}). Note, however,
that the gap edges remain symmetric with respect to the
Fermi energy, while the
spectral weight is larger below the Fermi level as
long as $t_{12}<t_{12,c}$. For
$t_{12}>t_{12,c}$, a rearrangement of
spectral weight from below to above the Fermi energy takes place,
while the size of the gap does not change noticeably.

We interpret this structure as a replica of the Kondo resonance
induced indirectly by the interactions between the even and odd channel. However,
adding or removing electrons in the odd-parity
orbital would at least break one ``Kondo
bond'' in the even-parity channel, i.e., cost an energy $\sim J_{cf}$, explaining the appearance and
size of the gap for $t_{12}=0$. Note that, in contrast to the
weak-coupling regime, this explanation connects the
gap to the spin rather than to the
charge degrees of freedom.
As $t_{12}$ increases, the gap becomes smaller 
because the direct exchange interaction $J_{\rm ex}$ between the impurity sites 
competes with the Kondo effect, effectively reducing the local
moment, which is screened by the band states, and hence the corresponding
spin gap. However, we find $\Delta > J_{cf} - J_{\rm ex}$, which would be the
naive expectation, because the formation of dynamic spin correlations
between the impurities due to $J_{\rm ex}$ introduces a further
contribution to $\Delta$. Thus, the gap will be always finite
and develops a minimum close to $t_{12,c}$, where $J_{cf}=J_{\rm ex}$. For
$t_{12}>t_{12,c}$, the impurities form a non-local spin singlet and the
spin gap will scale with $J_{\rm ex}$, i.e., increase again slowly
with increasing $t_{12}$.
An estimate for $t_{12}\ge t_{12,c}$ yields 
$\Delta \sim J_{\rm ex}= 4t_{12}^2/U \approx 0.26$, which is in rough
agreement with the numerical value $\Delta \approx 0.16$ obtained from Fig.~\ref{fig_u15}(c).

\subsubsection{intermediate-coupling regime}

\begin{figure}[t]
\begin{center}
\vspace{5pt}
\includegraphics[width=7.5cm,clip]{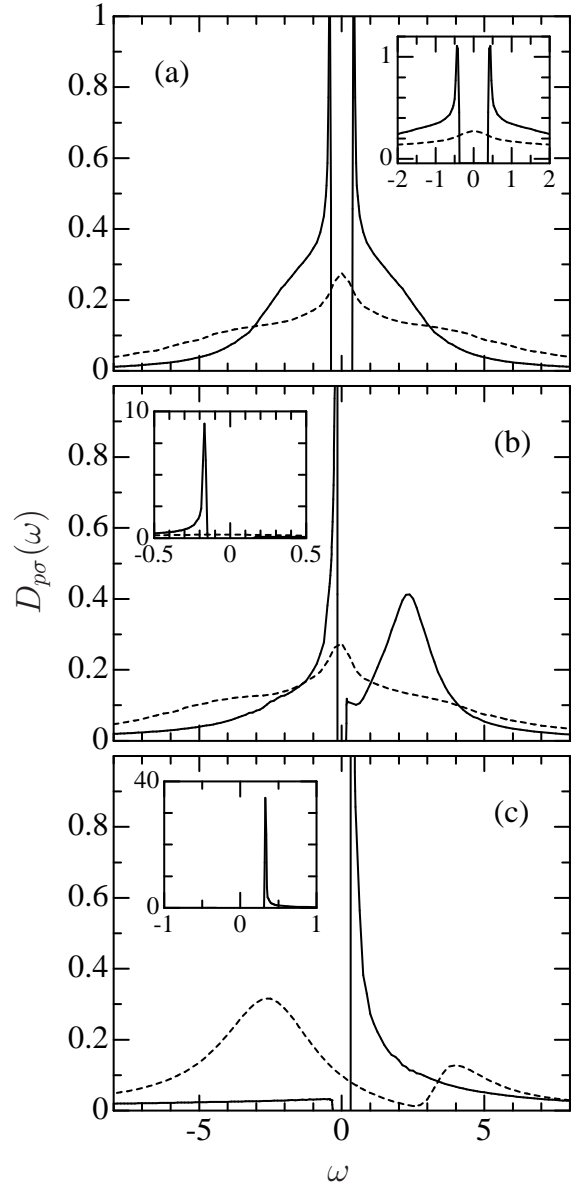}
\vspace{-5pt}
\end{center}
\caption{Spectral density of the coupled impurities for the odd-parity orbital 
$D_{o\sigma}(\omega)$ and for the even-parity orbital $D_{e\sigma}(\omega)$ 
with (a) $t_{12}=0$, (b) $t_{12}=0.5\pi\Delta_0$, and (c) $t_{12}=\pi\Delta_0$ at 
$U=5\pi\Delta_0$ and $W=20\pi\Delta_0$. Dashed lines denote 
$D_{e\sigma}(\omega)$ calculated with a constant discretization for $N=58$, 
$\Delta \varepsilon \approx 0.34\pi \Delta_0$, and $\eta=0.5\pi\Delta_0$, 
then deconvolved. Solid lines denote $D_{o\sigma}(\omega)$ calculated with a 
variable discretization for $N=118$, $0.01 \pi \le \Delta \varepsilon/\Delta_0 \le 1.45\pi$ 
and a constant broadening $\eta=0.02\pi\Delta_0$, then deconvolved. 
Insets: expanded view around the Fermi level $\omega=0$.}
\label{fig_u5}
\end{figure}

Finally, we present the spectral function for $U \approx
4\pi\Delta_0$, i.e., in the intermediate-coupling regime.
Note that from the point of view of the SIAM this value already resides
within the ``strong-coupling'' regime delimited by 
$\frac{U}{\pi\Delta_0}=2$ (the effective hybridization for the
even-parity channel is $2\Delta_0$).
In Fig.~\ref{fig_u5}, we show the spectral density calculated with 
the DDMRG for $t_{12}=0$, $0.5\pi\Delta_0$, and $\pi\Delta_0$ at 
$U=5\pi\Delta_0$. The transition here occurs at $t_{12,c}\sim 0.65\pi\Delta_0$. 
As already mentioned above, the even-parity spectral density 
$D_{e\sigma}(\omega)$ below $t_{12,c}$ resembles that of the SIAM. The
central peak at $\omega=0$ starts to become narrower and its spectral
weight is increasingly transferred to the high-energy range with increasing 
$U$. However, the three-peak structure typical of the
strong-coupling regime is not yet fully developed; the Hubbard bands
appear only as visible but shallow shoulders around $\omega\sim\pm
U/2$.
Moreover, in contrast to the SIAM, the central peak here does not
fully reach the Friedel limit, its height being slightly lower than
$1/\pi$. 
This behavior was also seen in previous NRG studies~\cite{sakai} and
in the mean-field 
approach~\cite{sire}, and is connected to the non-local magnetic
correlations induced by the RKKY exchange, which is stronger for
smaller $U$. As in the strong-coupling regime, $D_{e\sigma}(\omega)$ hardly changes 
with increasing $t_{12}$ until $t_{12,c}$ is reached, where a dramatic
redistribution of spectral weight connected to the formation
of a non-local singlet due to the direct exchange introduced by
$t_{12}$ appears. 
In particular, as in the strong-coupling regime, the Kondo
peak has vanished completely. 

The behavior of the odd-parity channel in Fig.~\ref{fig_u5} is also similar to the
strong-coupling limit. We again find a gap in the spectrum,
which remains symmetric about $\omega=0$ for all $t_{12}$, and observe
a similar but much more 
pronounced change in the distribution of spectral weight, as in
Fig.~\ref{fig_u15}. The gap, too, initially decreases until 
$t_{12,c}$ is reached and then increases again. We believe that the physics
behind this behavior is essentially the same as in the
$U\to\infty$ limit, although the energy scales associated with
the spin gaps induced by the different exchange mechanism in
particular now cannot
be written down explicitly. However, we expect them to be larger than
in the limit of large $U$, which is indeed what we observe in Fig.~\ref{fig_u5}.

\section{Conclusion}
In this paper, we have presented static and dynamical properties of the
two-impurity Anderson model at half filling for vanishing impurity
separation. In this limit, the otherwise rather complex model becomes
considerably simpler, nevertheless retaining most of its interesting physical
aspects. In particular, the competition between different types of
magnetic correlations such as Kondo, RKKY and superexchange is
preserved. We employ the (dynamical) density-matrix renormalization
method to calculate static properties (filling, magnetic correlations)
and one-particle spectra. For the latter, we obtain results that are
convoluted with a Lorenzian, which we, however, can deconvolute with
good accuracy, as demonstrated for the exactly solvable case $U=0$.

While this model can, in fact, easily be studied by e.g., Wilson's NRG,
an evident advantage of the DDMRG is that one obtains an accurate
description of spectral features on all energy scales at the cost of
the resolution of exponentially small structures. 
In the NRG, on the other hand, exponentially small scales
can be readily resolved, however, at the expense of accuracy at
intermediate and high energies. Since we are interested here in
describing both the features emerging at intermediate 
($\sim J_{\rm RKKY}$, $J_{\rm ex}$) and high energy scales (Hubbard bands
$\sim U/2$) as well as at a possible small Kondo scale, we feel that the DDMRG 
is more useful for the present study.

In the weak-coupling regime, $U<2\pi\Delta_0$, we observe a transition
between a situation with weakly ferromagnetically coupled impurity spins
to a situation with weak antiferromagnetic correlations 
as function of inter-impurity hopping $t_{12}$, as is apparent
from the behavior of $\langle \vec S_1\cdot\vec S_2\rangle$ in
Fig.~\ref{fig2}(c). From the spectral functions discussed in
Fig.~\ref{fig_u1}, we can furthermore infer that this transition is
primarily driven by the charge degrees of freedom in the odd-parity
channel of the two-impurity system. For $U>2\pi\Delta_0$, on the other
hand, the transition is into a state with rather strong
antiferromagnetic correlations. At the same time, the behavior of the
spectral functions, in particular in the odd-parity channel, changes
considerably. Although the transition is still accompanied by a change
in occupancy in the odd-parity channel, this change is visibly
reduced. Furthermore, the gap in the odd-parity spectrum is always
pinned to the Fermi energy and is of the size of the typical magnetic
exchange interactions, which points to the spin degrees of
freedom as the driving force of the transition. Finally, the
existence of a Kondo peak in the even-parity spectrum for
$t_{12}<t_{12,c}$, which abruptly vanishes for $t_{12}>t_{12,c}$, must be taken
as evidence that the transition is in fact driven by the competition
between Kondo and direct exchange in this regime. It is quite
important to note that the actual transition occurs at the point
where $J_{cf}=J_{\rm ex}$ and {\em not} when $T_K(J_{cf})=J_{\rm ex}$, and
also that the energy scales appearing in the spectra in Figs.~\ref{fig_u15}
and \ref{fig_u5} are in fact related to these ``intermediate''
quantities rather than the actually much smaller Kondo scale.

Of course, the current investigation is restricted to a special limit
of the TIAM, namely, that of  vanishing impurity
separation. The 
results for this case clearly show that a more thorough investigation
of this model is still necessary. 
In particular, we believe that such an investigation
must carefully study the relation between the different energy scales
inherent to the problem.
The method to study the TIAM should be
chosen so that it can at least resolve accurately intermediate,
i.e., of the order of the effective exchange interactions, and
low energy scales, i.e., of the order of the Kondo scale. 
Evidently, we cannot expect that any method can handle both
regimes equally well when the energy scale of the latter is
exponentially small, but we
believe that the DDMRG can at least treat the intermediate situation
when $T_K\ll J_{\rm ex}$ but is still resolvable with the method.

\ack

We thank F.~Gebhard and E.~Jeckelmann
for useful discussions.  
This work was supported in part by   
the Deutsch Forschungsgemeinschaft through the collaborative 
research center SFB 484.
Part of the computations were carried out at the Norddeutsche 
Verbund f\"ur Hoch- und H\"ochstleistungsrechnen.

\end{document}